\documentclass[aps,prab,twocolumn,amsmath,amssymb,groupedaddress]{revtex4-2}

\usepackage{siunitx}
\usepackage{placeins}
\usepackage{float}
\usepackage[colorlinks=true]{hyperref}
\hypersetup{ 
    citecolor=black,%
    filecolor=black,%
    linkcolor=black,%
    urlcolor=black 
}
\usepackage{graphicx}
\usepackage{dcolumn}
\usepackage{bm}
\usepackage{color}

\begin{document}

\title{Predicting the transverse emittance of space charge dominated beams using the phase advance scan technique and a fully connected neural network}

\author{F.~Mayet}
\email{frank.mayet@desy.de}
\affiliation{Deutsches Elektronen-Synchrotron DESY, Notkestr. 85, 22607 Hamburg, Germany}
\author{M.~Hachmann}
\affiliation{Deutsches Elektronen-Synchrotron DESY, Notkestr. 85, 22607 Hamburg, Germany}
\author{K.~Floettmann}
\affiliation{Deutsches Elektronen-Synchrotron DESY, Notkestr. 85, 22607 Hamburg, Germany}
\author{F.~Burkart}
\affiliation{Deutsches Elektronen-Synchrotron DESY, Notkestr. 85, 22607 Hamburg, Germany}
\author{H.~Dinter}
\affiliation{Deutsches Elektronen-Synchrotron DESY, Notkestr. 85, 22607 Hamburg, Germany}
\author{W.~Kuropka}
\affiliation{Deutsches Elektronen-Synchrotron DESY, Notkestr. 85, 22607 Hamburg, Germany}
\author{T.~Vinatier}
\affiliation{Deutsches Elektronen-Synchrotron DESY, Notkestr. 85, 22607 Hamburg, Germany}
\author{R.~Assmann}
\affiliation{Deutsches Elektronen-Synchrotron DESY, Notkestr. 85, 22607 Hamburg, Germany}

\date{\today}

\begin{abstract}
The transverse emittance of a charged particle beam is an important figure of merit for many accelerator applications, such as ultra-fast electron diffraction, free electron lasers and the operation of new compact accelerator concepts in general. One of the easiest to implement methods to determine the transverse emittance is the phase advance scan method using a focusing element and a screen. This method has been shown to work well in the thermal regime. In the space charge dominated laminar flow regime, however, the scheme becomes difficult to apply, because of the lack of a closed description of the beam envelope including space charge effects. Furthermore, certain mathematical, as well as beamline design criteria must be met in order to ensure accurate results. In this work we show that it is possible to analyze phase advance scan data using a fully connected neural network (FCNN), even in setups, which do not meet these criteria. In a simulation study, we evaluate the perfomance of the FCNN by comparing it to a traditional fit routine, based on the beam envelope equation. Subsequently, we use a pre-trained FCNN to evaluate measured phase advance scan data, which ultimately yields much better agreement with numerical simulations. To tackle the confirmation bias problem, we employ additional mask-based emittance measurement techniques.
\end{abstract}

\maketitle

\section{Introduction}
Many modern particle accelerators are tuned to achieve as small transverse beam emittance as possible. This is due to the fact that most users demand for the highest beam brightness possible. Beam brightness is important for many accelerator applications, such as ultra-fast electron diffraction \cite{doi:10.1098/rsta.2005.1735}, free electron lasers \cite{RevModPhys.88.015006} and the operation of new compact accelerator concepts in general (e.g. \cite{England:2014bf,KARTNER201624,Giorgianni:2016ds,Nghiem:437316}). A common definition of brightness is \cite{doi:10.1063/1.860076}
\begin{equation}
	B = \frac{\eta I}{\pi ^2 \varepsilon_x \varepsilon _y},
\end{equation}
where $\eta$ is a form factor close to unity, $I$ is the beam peak current and $\varepsilon_{x,y}$ the horizontal and vertical transverse emittance respectively. Hence, in order to maximize $B$, transverse emittance has to be minimal. 

There are multiple methods to characterize the transverse emittance. One of the most common techniques is the phase advance scan technique, where the transverse beam size is recorded on a screen vs.\ the focusing strength of an upstream quadrupole or solenoid magnet \cite{ReesRivkin:slac1984,Minty:2003,HACHMANN2016318}. The data can then be fitted based on the beam envelope equation. Space charge effects can be included to some extent \cite{Hachmann:138874,PhysRevAccelBeams.20.013401}. Instead of scanning the focusing strength of a magnet, also multiple screens can be used to record the beam size vs.\ the phase advance. Other -- potentially single-shot -- methods involve insertion of masks into the beamline, which then, subsequently, can be imaged on a downstream screen \cite{Zhang:fermilab1996}. Coupled with advanced reconstruction algorithms these methods are capable of delivering reconstruction of the core 4D phase space \cite{PhysRevAccelBeams.21.102802}. In this work we concentrate on the phase advance scan technique, as this is the easiest one to implement, only requiring standard beamline components.

One of the limitations of the phase advance scan technique is that there is no closed description of the beam envelope for space charge dominated beams \cite{Hachmann:138874,PhysRevAccelBeams.20.013401}. It is therefore difficult to apply the method in this regime. Space charge dominated beams especially occur, for example, in the injector part of high-brightness electron sources, where the beam is still non-relativistic. In order to quantify whether a beam is space charge dominated, the so-called \emph{laminarity parameter} $\rho$ can be calculated \cite{Ferrario:1982426}. This parameter represents the ratio between the space charge term and the emittance term of the beam envelope equation. It is given by
\begin{equation}
 	\rho = \frac{I\sigma^2}{2I_\text{A}\gamma \varepsilon _n^2},
 	\label{eq:laminarity}
 \end{equation}
where $I$ is the peak current of the beam, $I_\text{A} \approx \SI{17}{\kilo\ampere}$ is the Alfvén current and $\varepsilon _n = \beta \gamma \varepsilon$ is the normalized emittance with the Lorentz factor $\gamma$ and $\beta = v/c$. In case $\rho \gg 1$, the beam can be considered as space charge dominated (laminar flow regime). Otherwise the evolution of the beam envelope is dominated by the emittance pressure (thermal regime).

In this work we show in simulation that it is possible to successfully analyse phase advance scan data for $\rho \gg 1$ beams using a pre-trained fully connected neural network (FCNN). Subsequently, we apply the method to measured data. Machine learning and neural networks in particular have recently been used in the context of accelerators for various purposes. These include, among others, fault detection of machine components \cite{Solopova:IPAC2019-TUXXPLM2}, machine stability optimization and analysis \cite{PhysRevX.10.031039,PhysRevLett.121.044801,PhysRevAccelBeams.23.074601}, virtual diagnostics \cite{Sanchez-Gonzalez2017,PhysRevAccelBeams.21.112802,Ratner:21}, beam quality optimization in plasma accelerators \cite{PhysRevLett.126.174801,PhysRevLett.126.104801} and orders of magnitude speed-up in multiobjective optimization of accelerator parameters \cite{PhysRevAccelBeams.23.044601}. Here, we focus on the analysis of otherwise difficult to interpret measurement data.
\section{Measurement Technique} 
\label{sec:MeasurementTechnique}
The state of a single particle with respect to a given design, or reference trajectory, is usually defined by the 6D phase space vector
\begin{equation}
	\mathbf{X} = (x, x^\prime, y, y^\prime, z, \delta)^\text{T},
\end{equation}
where $x^\prime = p_x/p_z$ and $y^\prime=p_y/p_z$ are the horizontal and vertical divergence respectively, $x,y,z$ the distances of the particle from the reference trajectory and $\delta = \Delta p / p_0$ is the relative deviation of the particle's individual momentum from the reference momentum. $p_x, p_y, p_z$ are the three momentum components and $()^\text{T}$ denotes the transpose of a matrix. We are interested in the evolution of the 4D transverse phase space vector
\begin{equation}
	 \mathbf{X^\text{(tr)}} = (x, x^\prime, y, y^\prime)^\text{T}.
\end{equation}
Assuming negligible correlation between the evolution of the phase space coordinates in the $x$ and $y$ planes, it is possible to treat them separately, yielding the two sub-space vectors
\begin{equation}
	\mathbf{x} = (x, x^\prime)^\text{T}, \hspace{0.5cm}\mathbf{y} = (y, y^\prime)^\text{T}.
\end{equation}
A common framework to describe the evolution of a charged particle is linear beam optics, where only linear transformations of $\mathbf{x}$ and $\mathbf{y}$ are taken into account. Each beamline element is represented by a so-called transfer matrix defined by the relation
\begin{equation} \label{eq:matrixtransformation}
	\mathbf{x} = \mathbf{M}\cdot\mathbf{x_0},
\end{equation}
where $\mathbf{x_0} = (x_0, x^\prime_0)^\text{T}$ is the initial phase space coordinate and
\begin{equation}
	\mathbf{M} = 
	\begin{pmatrix}
		M_{11} & M_{12}\\
		M_{21} & M_{22}
	\end{pmatrix}.
\end{equation}
The $2\times2$ matrix for a simple drift is given by
\begin{equation}
	\mathbf{M}_\text{D}(s) = 
	\begin{pmatrix}
		1 & s\\
		0 & 1
	\end{pmatrix},
\end{equation}
where $s$ is the drift distance. Applying this matrix to $\mathbf{x_0}$ would result in $x = x_0 + x^\prime s, x^\prime = x^\prime_0$, as expected.

In an experiment only the rms beam size $\sigma_x = \sqrt{\langle x^2 \rangle}$ is accessible, where $\langle \rangle$ denotes the second central moment. Applying this to the general equation
\begin{equation}
	x = M_{11}x_0 + M_{12}x_0^\prime
\end{equation}
yields
\begin{equation}
	\begin{split}
	\sigma_x^2 = &M_{11}^2 \sigma_{x,0}^2 \\
	&+ 2M_{11}M_{12}\sigma_{x,0}(\sigma_{x,0})^\prime \\
	&+ M_{12}^2 \left( \frac{\varepsilon _x^2}{\sigma_{x,0}^2} + (\sigma_{x,0})^{\prime 2} \right)
	\end{split}
\label{eq:envelope}
\end{equation}
with $\varepsilon_x = \sqrt{\langle x^2 \rangle \langle x^{\prime 2} \rangle - \langle xx^\prime \rangle^2}$. Equation~\ref{eq:envelope} is the so-called rms envelope equation, which can be used to determine the transverse rms emittance $\varepsilon _x$ using a suitable (i.e. tunable) beam transformation $\mathbf{M}$. Note that Eq.~\ref{eq:envelope} does not take any space charge effects into account and is hence only valid in the $\rho \approx 1$ regime.

\begin{figure}[htbp]
  \centering
  \includegraphics[width=0.9\columnwidth]{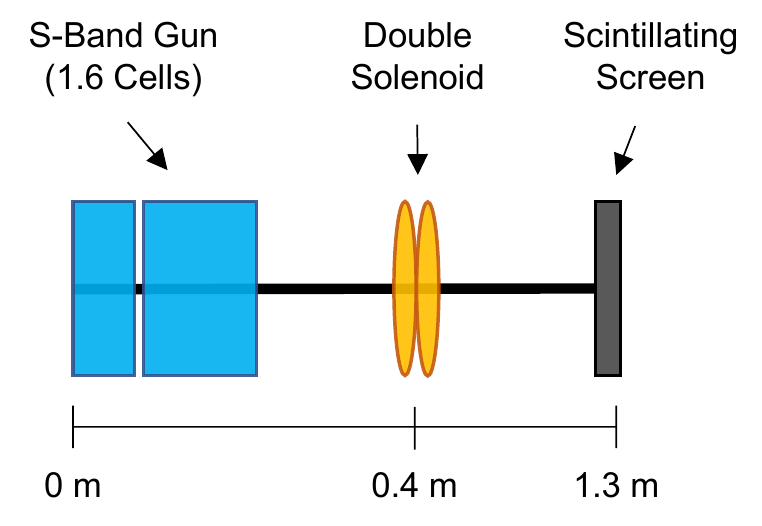}
  \caption{Schematic of a suitable beamline layout for transverse emittance measurements using the phase advance scan technique. The distances are based on what is installed at the ARES linac at DESY, Hamburg.}
  \label{fig:ARES_sketch}
\end{figure}

Figure~\ref{fig:ARES_sketch} shows a sketch of a potential measurement scenario. The elements and distances are chosen according to what is installed at the ARES electron linac at DESY, Hamburg \cite{instruments5030028}. The transfer matrix of the double solenoid magnet can be written as
\begin{equation}
	\begin{split}
		\mathbf{M}_\text{DS} &= \mathbf{M}_\text{TL} \cdot \mathbf{M}_\text{D}(l_\text{D}) \cdot \mathbf{M}_\text{TL} \\
		&=
		\begin{pmatrix}
			1 & 0 \\
			-\frac{1}{f} & 1
		\end{pmatrix} 
		\cdot
		\begin{pmatrix}
			1 & l_\text{D}\\
			0 & 1
		\end{pmatrix}
		\cdot
		 \begin{pmatrix}
			1 & 0 \\
			-\frac{1}{f} & 1
		\end{pmatrix}\\
		&=
		\begin{pmatrix}
			1-l_\text{D}/f & l_\text{D} \\
			(l_\text{D} - 2f)/f^2 & 1-l_\text{D}/f
		\end{pmatrix}, 
	\end{split}
	\label{eq:double_solenoid_matrix}
\end{equation}
where $l_\text{D}$ is the drift distance between the two single solenoids and $f$ the focal length of each solenoid. Here the approximation that $f$ is larger than the length of the solenoid was used, i.e. the thin lens approximation. The focal length of a solenoid is given by \cite{doi:10.1119/1.3129242}
\begin{equation}
	f(B_{z,\text{max}}) = \left[ \left( \frac{q}{2  \overline p_z } \right)^2 F_2\right]^{-1}, 	
\end{equation}
where $B_{z,\text{max}}$ is the peak magnetic field, $q$ the particle charge and $\overline p_z$ the average longitudinal beam momentum. $F_2 = \int B_z^2 \text{d}z \propto B_{z,\text{max}}^2$ is the second field integral of the on-axis magnetic field. By inserting the expression $\mathbf{M}_\text{D}(l_\text{S}) \cdot \mathbf{M}_\text{DS}$, where $l_\text{S}$ is the drift between the solenoid and the screen, in Eq.~\ref{eq:envelope}, it can be seen that now the $M_{ij}$ elements can be conveniently adjusted in the experiment as $B_{z,\text{max}}$ is varied. The emittance at the position of the solenoid can thus be determined by fitting the recorded $\sigma_{x,i}$ vs.\ $B_{z,\text{max},i}$ at the screen with Eq.~\ref{eq:envelope}.

It is possible to include transverse space charge forces into the model to some extent. This is done by including a defocusing term in the drift between the focusing element and the screen. Considering a uniformly charged cylindrical bunch with radius $R$ and length $L$, the envelope equation then reads in differential form \cite{PhysRevAccelBeams.20.013401}
\begin{equation}
	\sigma _x ^{\prime \prime} - \frac{P}{4\sigma_x} G(\xi, A) - \frac{\varepsilon _x ^2}{\sigma_x ^3} = 0,
	\label{eq:envelope_diff_eq_spch}
\end{equation}
where $G(\xi, A) \in [0,1]$ is a form factor, which depends on the centered longitudinal intra-bunch coordinate $\xi$ and the rest frame aspect ratio $A=R/(\gamma L)$. $P$ is the so-called generalized perveance given by
\begin{equation}
	P = \frac{eQ}{2 \pi \epsilon_0 m_e c^2 L \beta ^2\gamma^3},
\end{equation}
where $Q$ is the total charge of the bunch, $\epsilon _0$ the vacuum permittivity and $m_e$ the electron rest mass. In principle Eq.~\ref{eq:envelope_diff_eq_spch} can now be used to construct a similar fit model to Eq.~\ref{eq:envelope}. There are a number of caveats to take into account, however:
\begin{itemize}
	\item The model is only fully valid for a perfectly cylindrical bunch.
	\item The form factor $G$ depends on the aspect ratio, which depends on the transverse beam size and bunch length, both not being constant in the experiment (especially in the over-focused part of the scan).
	\item Since the concept of the envelope equation relies on the emittance being a constant of motion, non-linear space charge forces are intrinsically neglected in this approach.
	\item The perveance term depends on the bunch length, which might not be accessible to sufficient precision in the experiment.
	\item Equation~\ref{eq:envelope_diff_eq_spch} cannot be solved analytically and has to be approximated by a polynomial series (see \cite{Hachmann:138874} for a detailed description).
\end{itemize}
All of these caveats lead to the conclusion, that the envelope equation based data analysis method is not ideal in the $\rho \gg 1$ regime. 

Independent of the value of $\rho$, two more criteria need to be met in order to ensure an accurate fit result. Considering the third term of Eq.~\ref{eq:envelope}, the purely mathematical criterion
\begin{equation}
	\frac{\varepsilon _x ^2}{\sigma _{x,0}^2 \cdot (\sigma _{x,0})^{\prime 2}} \geq 0.01
	\label{eq:fit_criterion_math}
\end{equation}
can be derived \cite{Hachmann:138874}. This criterion ensures the numerical significance of $\varepsilon _x$. The second criterion is based on the fact that the scan needs to include a minimum. The initial beam optics needs to be setup, such that a potential focus of the beam lies behind the screen used to measure the beamsize. At the same time, the focusing element needs to be strong enough to focus the beam onto the screen, which implies a constraint on the distance between focusing element and screen. Using Eq.~\ref{eq:envelope}, these considerations can be summarized by the criterion
\begin{equation}
	-\frac{\sigma_{x,0} \cdot (\sigma_\text{eff})^\prime}{f_\text{max}^2 \cdot \left((\sigma_\text{eff})^{\prime 2} + \frac{\varepsilon_x ^2}{\sigma_{x,0}^2}\right)} \leq l_\text{S} \leq \frac{\sigma _{x,0}^2}{2 \varepsilon_x},
\end{equation}
where $(\sigma_\text{eff})^{\prime} = - (\sigma _{x,0}/f_\text{max} - (\sigma _{x,0})^\prime)$. In case both of the two criteria are fulfilled, the emittance can be retrieved.

Based on the aforementioned considerations, we propose using an alternative way to analyze phase advance scan data. Specifically, we propose using a pre-trained FCNN to overcome the problem of the incomplete fit model in $\rho \gg 1$ cases, as well as the criterion described by Eq.~\ref{eq:fit_criterion_math}. To this end, we have performed a simulation study, which is presented in detail in the following sections. The resulting FCNN was then subsequently applied to real world data, as shown below.
\section{Simulation Study - Methodology}
\label{sec:Methodology}
It has been shown already in 1989 that neural networks with only one unbounded hidden layer can approximate any Borel measureable function from finite dimensional space to another to arbitrary precision \cite{Cybenko1989,HORNIK1989359}. More recent research focuses on the expressiveness (approximation accuracy) of both depth (i.e. the number of hidden layers) and width (i.e. the number of artificial neurons in a layer) bounded networks. In \cite{NIPS2017_32cbf687}, for example, the authors show that any Lebesgue integrable function $f: \mathbb{R}^n \rightarrow \mathbb{R}$ on $n$-dimensional space can be approximated to arbitrary accuracy by a fully connected width-$(d_\text{in}+4)$ ReLU network with respect to the $\ell_1$ norm as a measure of approximation quality. In other words, the network represented by the transfer function $F$ satisfies $\int _{\mathbb{R}^n} |f(x) - F(x) | \text{d}x < \epsilon, \forall \epsilon > 0$ (see \cite{NIPS2017_32cbf687}, Theorem 1). ReLU here refers to the the so-called Rectified Linear Unit neuron activation function, definded by $\text{ReLU}(x) = \text{max(0,x)}$ and $d_\text{in}$ is the input dimensionality. This width boundary $w$ has since been refined and generalized for example in \cite{hanin2018approximating,Hanin_2019} to be $d_\text{in} + 1 \leq w_\text{min}(d_\text{in},d_\text{out}) \leq d_\text{in} + d_\text{out}$, for functions of the form $f:[0,1]^{d_\text{in}} \rightarrow \mathbb{R}^{d_\text{out}}$, where $d_\text{in}$ and $d_\text{out}$ are the input and output dimensionality respectively. Limits on the depth can be estimated in specific cases in terms of the so-called modulus of continuity of $f$, given by $\omega_f(\varepsilon) = \text{sup}\{|f(x)-f(y)|||x-y|\leq \varepsilon\}$, where $\varepsilon$ is an arbitrarily small change in the argument of $f$. For continuous functions $f:[0,1]^{d_\text{in}} \rightarrow \mathbb{R}_+$ the depth of a $d_\text{in}+2$ wide network $\mathcal N$ can, for example, be expressed as $\text{depth}(\mathcal N _\varepsilon) = 2\cdot d_\text{in}! / \omega_f(\varepsilon)^{d_\text{in}}$, cf. \cite{hanin2018approximating}. We note that in practice the specific layout of a neural network is often determined experimentally, as the aforementioned boundaries are merely based on proofs of existence.

In this study, we aim to map the phase advance scan data to the normalized transverse emittance at the focusing element. Mathematically, this means we assume a connection of the scan data to the physical quantity of the form $f:\mathbb{R}_+^{d_\text{in}} \rightarrow \mathbb{R}_+$, where the dimensionality $d_\text{in}$ is given by the number of scan data points. Note that, based on the knowledge of the problem, we can always map (normalize) the input data from $\mathbb{R}_+^{d_\text{in}}$ into $[0,1]^{d_\text{in}}$. The function $f$ operates on the measure space $(\mathbb{R}_+^{d_\text{in}}, \mathcal B, \lambda)$ with the Borel-$\sigma$-algebra $\mathcal B$ and the Lebesgue measure $\lambda$. It is hence measureable in the mathematical sense. In addition, we expect $f$ to be a continuous function based on the physical background of the problem. We can hence conclude that $f$ is Lebesgue integrable and suitable to be approximated for example by a width bounded ReLU network. 

\begin{figure}[htbp]
  \centering
  \includegraphics[width=0.72\columnwidth]{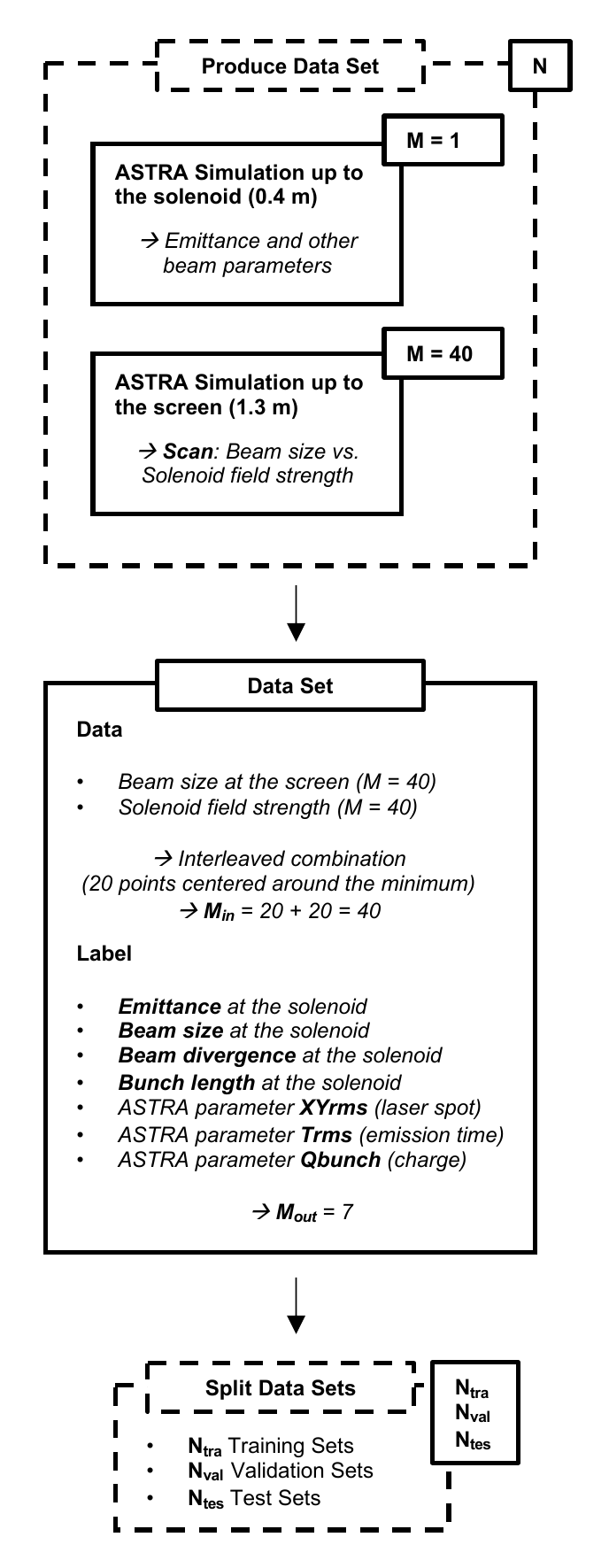}
  \caption{Diagram showing how the FCNN training, validation and test data is produced. $N$ is the total number of data sets. $M$ is the number of ASTRA simulations. $M_\text{in}$ is the final number of data points to be fed to the FCNN. $M_\text{out}$ is the length of the output vector, i.e. the number of predicted beam and simulation parameters. The data sets are split into three categories, where in this study the common split of $N_\text{tra} = 0.6\cdot N$, $N_\text{val} = 0.2\cdot N$ and $N_\text{tes} = 0.2\cdot N$ is used.}
  \label{fig:Methodology_1}
\end{figure}

To validate this approach, we setup a simulation study based on the simple beamline layout shown in Fig.~\ref{fig:ARES_sketch}. The main simulation study is split into three parts:
\begin{enumerate}
	\item Building a large number of data sets (training, validation and test),
	\item Training the FCNN and evaluation of the performance using the test data sets,
	\item Comparison of the FCNN performance to the traditional fit method, as discussed above.
\end{enumerate}
The first step of the simulation study is to build a large number of data sets. Creating a data set consists of two steps:
\begin{itemize}
	\item Numerical tracking of the beam from the cathode to the location of the solenoid,
	\item Numerical simulation of the solenoid scan.
\end{itemize}
First, the emittance at the solenoid position is determined by numerical tracking of the particles. In this step the solenoid field is set to zero. In addition to the emittance, other beam parameters, such as the beam size, divergence, or bunch length can be recorded as well. Then, the simulation domain is extended up to the position of the screen, which is used in the experiment to record the beam size vs.\ the solenoid focusing strength. The experiment is then simulated for $M$ focusing strength settings. It is important to setup the scan range such that the resulting data includes the beam size minimum, i.e. the focus, as it carries most of the information about the emittance at the solenoid position \cite{Hachmann:138874}. We use the well established code \textsc{ASTRA} \cite{ASTRAASpaceChar:LTSRiAsm}, which takes space charge effects into account. The beam size vs.\ focusing strength scan data functions as the data set to be interpreted by the FCNN. Over the course of the study, a specific way to prepare the input data turned out to yield the best results. For each scan, $M/2$ scan points centered around the minimum beam size are interleaved with the relative focusing strength difference $\delta B_i = (B_i - B_\text{foc})/B_\text{foc}$, where $B_\text{foc}$ is the setting corresponding to the minimal beam size. The data set $S_\text{in}$ is then of the form
\begin{equation}
	S_\text{in} = [\delta B_1, \sigma_1, \delta B_2, \sigma_2, ..., \delta B_{(M/2)}, \sigma_{(M/2)}],
	\label{eq:data_set_form}
\end{equation}
where $\sigma_i$ is the $i$th rms beam size. Each of these data sets is labeled with a set of important beam and simulation input parameters. These labels are then used to perform so-called supervised training of the FCNN. After the learning process, the FCNN is able to predict each of these parameters from given scan data, which is prepared according to Eq.~\ref{eq:data_set_form}. Figure.~\ref{fig:Methodology_1} summarizes how the data sets are created and what types of data they contain. 

For this particular study $N = 16066$ random data sets with $M=40$ were produced. Each data set differs in the three key \textsc{ASTRA} input parameters \emph{total charge}, \emph{laser spot size} and \emph{cathode emission time}. Table~\ref{tab:data_set_ranges} summarizes the parameter ranges used for this study, which are losely based on typical settings at the ARES linac at the time. The parameters are varied according to a uniform distribution. The ARES S-band gun was simulated with an, at the time available, peak gradient of \SI{65}{\mega \volt/\meter}, resulting in a final $\gamma = 6.8$. Based on the parameter ranges shown in Table~\ref{tab:data_set_ranges}, a convergence study in terms of required macro particles in the numerical simulation was performed. For the highest possible charge density, 10000 particles were found to be sufficient.

\begin{table}[htbp]
   \centering
   \caption{\textsc{ASTRA} input parameter ranges for the data set production.}
   \begin{ruledtabular}
   \begin{tabular}{lc}
   Parameter & Value \\
   \hline
   Bunch charge & $[0.01, 2.1]\,\SI{}{\pico\coulomb}$\\
   Laser spot size (flat top diameter) & $[240, 400]\,\SI{}{\micro\meter}$\\
   Cathode emission time (rms) & $[60, 100]\,\SI{}{\femto\second}$\\
   \end{tabular}
   \end{ruledtabular}
   \label{tab:data_set_ranges}
\end{table}

The neural network was implemented using the TensorFlow framework \cite{tensorflow2015-whitepaper}. The input layer has $M$ neurons, corresponding to the length of $S_\text{in}$. Then one hidden layer with $M$ and two hidden layers with $M/2$ neurons are added in order to capture non-linearities in the system. The system is then coupled to the output layer of size $M_\text{out}$, which corresponds to the number of \textsc{ASTRA} input and simulated beam parameters to be predicted by the network. The overall layout is hence
\begin{equation}
	M \rightarrow [M]-[M/2]-[M/2] \rightarrow M_\text{out}.
\end{equation}
This particular layout was determined empirically. Each neuron is coupled to every neuron of the following layer, or in other words the layers are fully connected. The neurons are activated using the well established rectified linear activation function (ReLU) \cite{pmlr-v15-glorot11a,ramachandran2017searching}. Training of the network is performed using a combination of the \textsc{adam} and \textsc{adagrad} gradient decent algorithm \cite{ruder2017overview} with the mean squared error (\textsc{MSE}) as the loss function. The available $N$ data sets are split into three categories. $N_\text{tra}$ training sets, $N_\text{val}$ validation sets and $N_\text{tes}$ test sets. The training sets are used to adjust the neuron weights during the training procedure, while the performance of the network is judged after each so-called epoch based on the validation sets, which are not used during training. This is done to avoid overfitting the training data. An epoch refers to one forward and backward pass of the entire training data. Finally, the performance of the resulting model is determined using the test sets, which have not been part of the learning procedure at all. We use the common split of $N_\text{tra} = 0.6\cdot N$, $N_\text{val} = 0.2\cdot N$ and $N_\text{tes} = 0.2\cdot N$. The network was trained for $\sim 10000$ epochs using \textsc{adam} and another $\sim 10000$ epochs using \textsc{adagrad} \footnote{In our study, the training procedure usually took $<\SI{1}{\hour}$ on an Apple M1 processor. Production of the training data sets, however, can take several days, depending on the available compute infrastructure. In order to speed-up data set production, we used the parallelized version of ASTRA}.
\section{Simulation Study - Data Set}
\begin{figure}[b]
  \centering
  \includegraphics[width=\columnwidth]{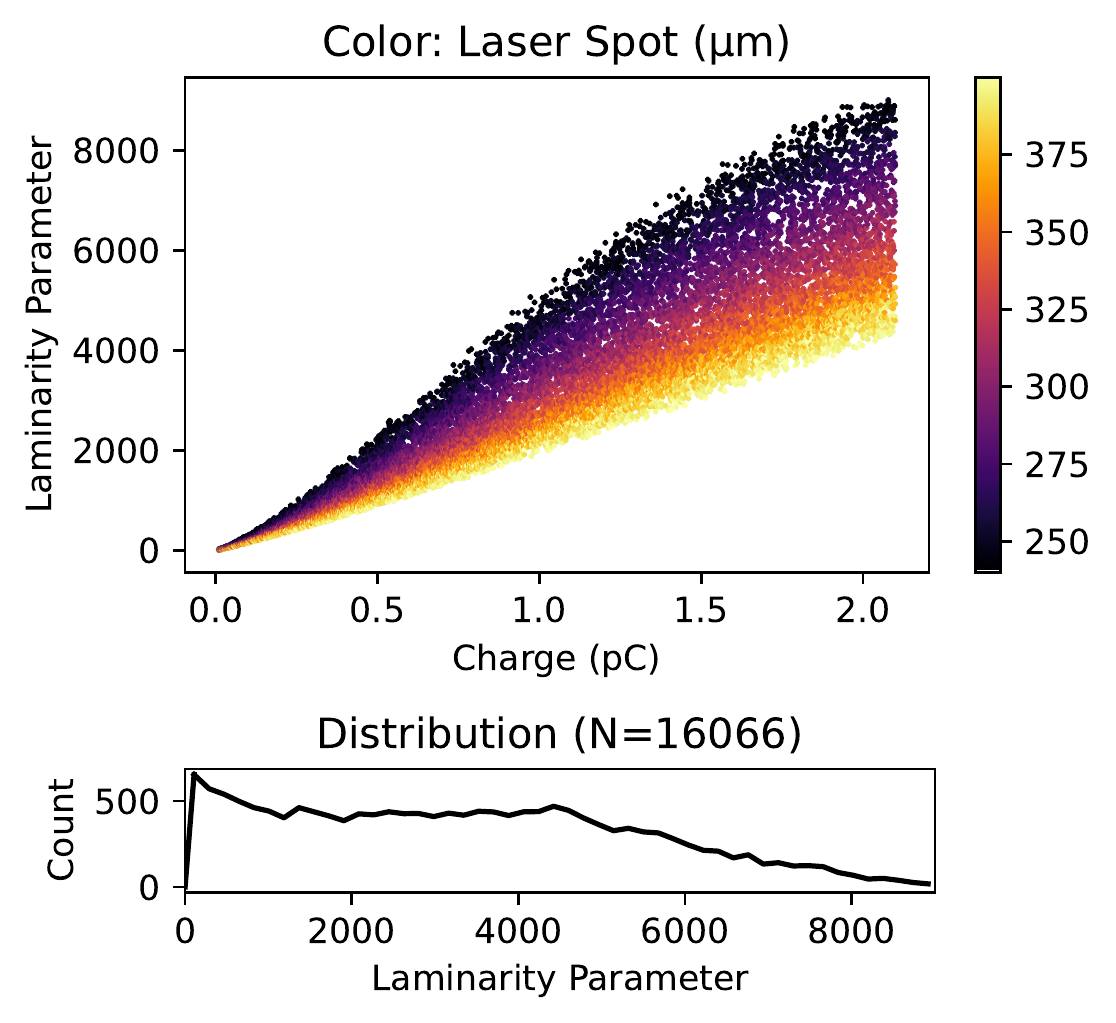}
  \caption{Top: Scatter plot of the laminarity parameter $\rho$ at the solenoid according to Eq.~\ref{eq:laminarity} vs.\ the bunch charge for all training data sets. $\gamma = 6.8$ for this population. The color indicates the laser spot size on the cathode for each data set. Bottom: Distribution of the laminarity parameter.}
  \label{fig:space_charge_dominance_train_data_sets}
\end{figure}

Before evaluating the prediction performance of the neural network, it is useful to inspect the training data set. Since we are especially interested in analyzing phase advance scan data for space charge dominated beams, the laminarity parameter $\rho$ at the solenoid position was calculated for each data set (cf. Eq.~\ref{eq:laminarity}). Figure~\ref{fig:space_charge_dominance_train_data_sets} shows $\rho$ vs.\ the bunch charge. The color scale indicates the laser spot size on the cathode used in the particular simulation. In addition, the distribution of $\rho$ across the whole data set is shown.
It can be seen that all of the data sets lie in the $\rho \gg 1$, i.e. space charge dominated, regime ($\rho_\text{min} = 17.8$). Also, the higher the charge, the higher the value for $\rho$, as expected. In addition, the color scale reveals that the smaller the laser spot size on the cathode, the higher the value for $\rho$. The sensitivity of $\rho$ on the laser spot size strongly depends on the bunch charge.

\begin{figure}[htbp]
  \centering
  \includegraphics[width=\columnwidth]{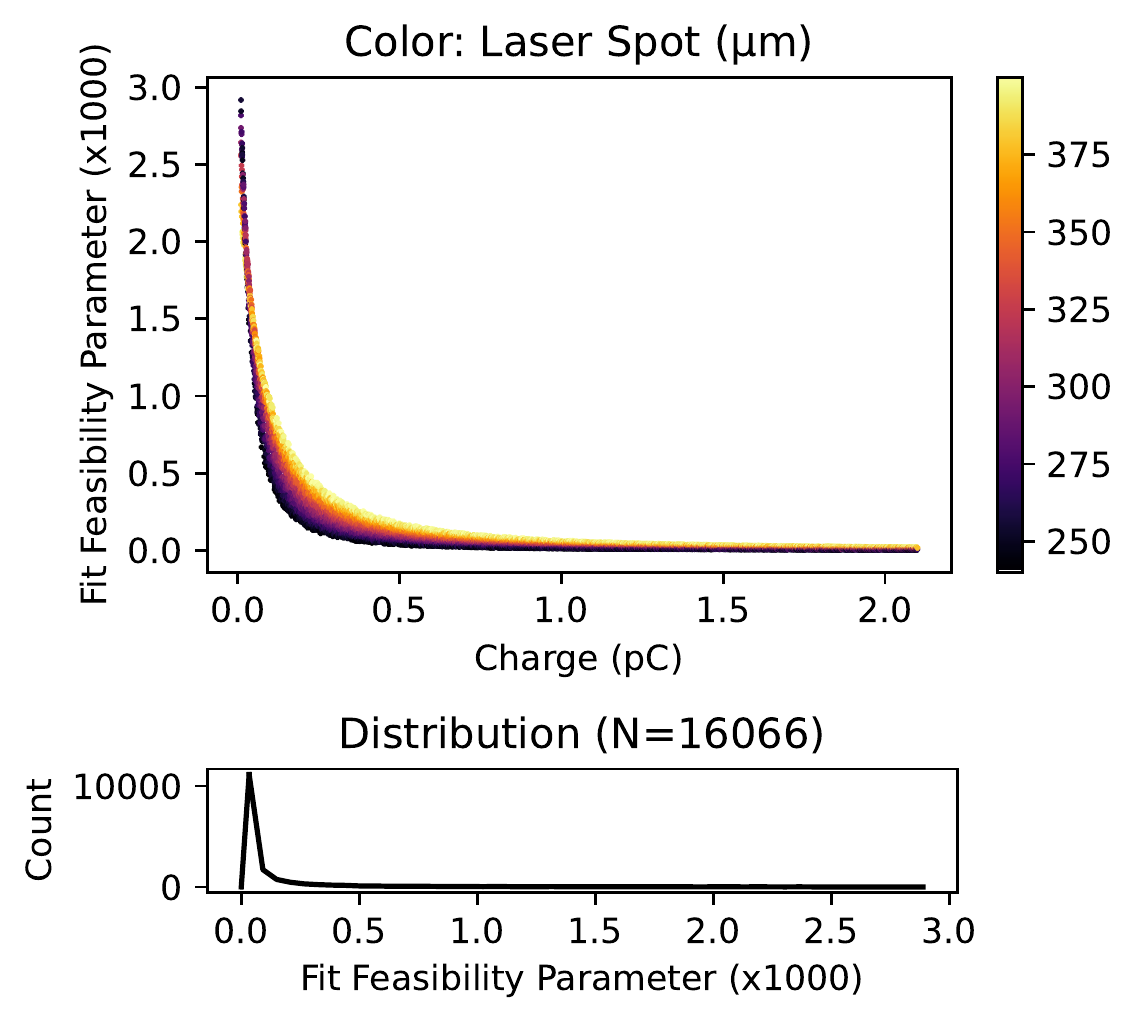}
  \caption{Top: Scatter plot of the fit feasibility parameter according to Eq.~\ref{eq:fit_criterion_math} vs.\ the bunch charge for all training data sets. $\gamma = 6.8$ for this population. The color indicates the laser spot size on the cathode for each data set. Bottom: Distribution of the fit feasibility parameter. \emph{Note that the plot shows the values multiplied by 1000 for readability.}}
  \label{fig:feasibility_train_data_sets}
\end{figure}

As noted above, the traditional fit method only works if Eq.~\ref{eq:fit_criterion_math} is satisfied. Figure~\ref{fig:feasibility_train_data_sets} shows the fit feasibility criterion for each data set, with the same color code as in Fig.~\ref{fig:space_charge_dominance_train_data_sets}. None of the data sets satisfies the criterion, which leads to the expectation that the traditional fit method should not work well on the training data (and with that in reality for the ARES working point, which is the basis for the parameter space shown in Table~\ref{tab:data_set_ranges}).
\section{Results and Comparison} 
\label{sec:ResultsAndComparison}
\begin{figure*}[htbp]
  \centering
  \includegraphics[width=0.98\textwidth]{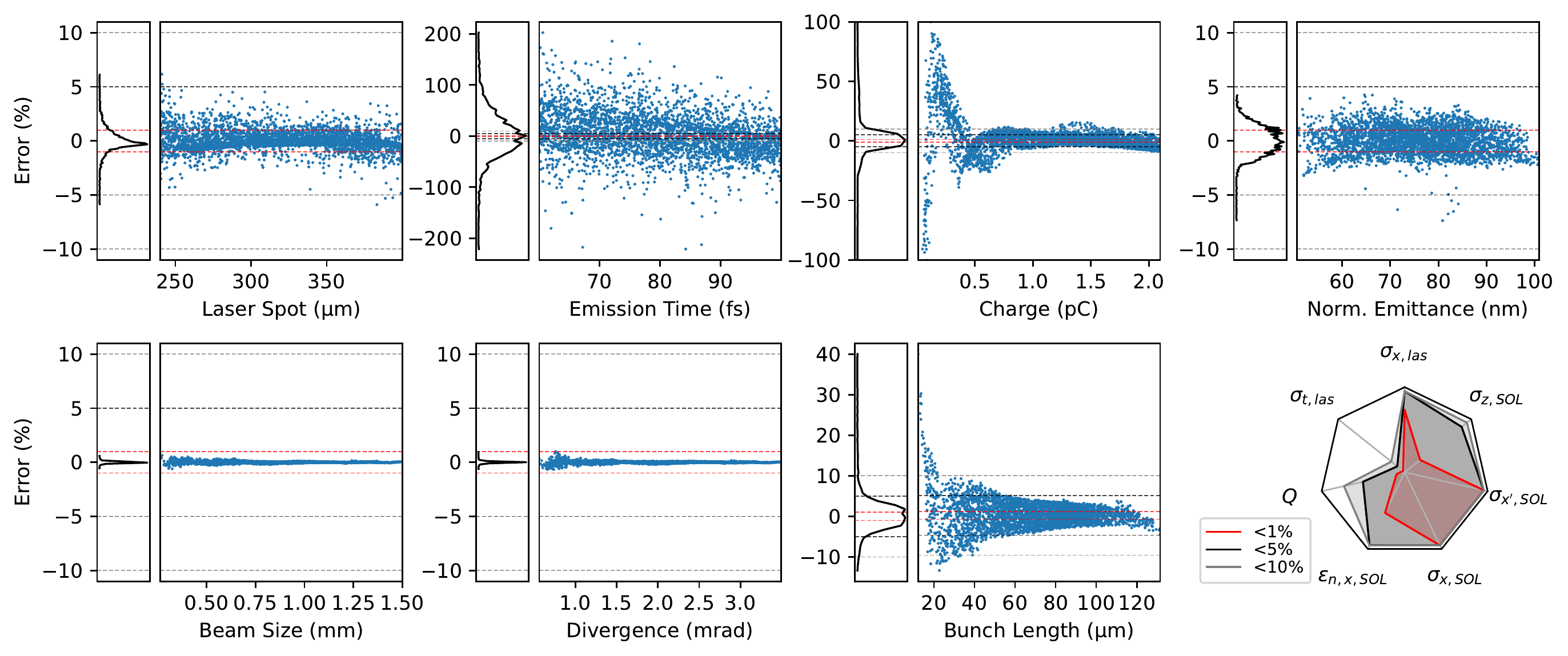}
  \caption{Relative error distributions for all verification data sets and label components. The last plot is a radar plot showing the percentage of data sets within a \SI{1}{\percent} (red), \SI{5}{\percent} (black) and \SI{10}{\percent} (gray) relative error interval respectively. The dashed lines in the distribution plots refer to the same error intervals. $Q:$ Bunch charge, $\sigma_{t,\text{las}}:$ Emission time, $\sigma_{x,\text{las}}:$ Laser spot size, $\sigma_{z,\text{SOL}}:$ Bunch length at the solenoid, $\sigma_{x^\prime,\text{SOL}}:$ Divergence at the solenoid, $\sigma_{x,\text{SOL}}:$ Beam size at the solenoid, $\varepsilon_{\text{n},x,\text{SOL}}:$ Normalized emittance at the solenoid.}
  \label{fig:error_dists_perf}
\end{figure*}

In this section the performance of the pre-trained FCNN is presented. We also compare its performance against the traditional fit method discussed above. In order to evaluate the performance of the FCNN, we try to predict the labels of the $N_\text{tes}$ test data sets, which were not used in the supervised training procedure. The main goal of the study is to predict the transverse emittance, but since the labels include a number of other simulation and beam parameters, the FCNN also provides predictions of these. In order to better quantify the prediction performance for the different label components, the relative error between predicition and truth was calculated for each data set. This is shown in Fig.~\ref{fig:error_dists_perf}. In addition to the error distributions, a radar plot visualizes the prediction performance in terms of number of data sets in \SI{1}{\percent}, \SI{5}{\percent} and \SI{10}{\percent} relative error intervals respectively (see Table~\ref{tab:pred_perf_table} for the actual percentages). In the ideal case, the heptagon would be filled completely. Inspection of the results reveals that some quantities are predicted much better than others. Specifically, it can be seen that the cathode emission time is predicted particularly bad. This result is somewhat expected, however, because this quantity refers to the longitudinal phase space at emission time, which cannot directly be accessed via a transverse beam size measurement. All quantities, which refer to the transvere phase space at the solenoid show very good prediction performance with $<\SI{5}{\percent}$ error. The prediction performance of both bunch charge and bunch length at the solenoid needs to be considered in more detail. In case of the bunch charge, values $\lesssim\SI{0.5}{\pico\coulomb}$ are predicted much less accurately. This can be explained by the lack of significant space charge effects, which alter the shape of the beam size vs.\ focusing strength curve, effectively leading to degeneracy w.r.t. the initial bunch charge. Despite being a quantity of the longitudinal phase space, the bunch length is generally predicted with an error $<\SI{10}{\percent}$. This is because here the bunch length is directly correlated with the bunch charge and hence space charge effects. The prediction performance decreases towards smaller bunch lengths, which can be explained by the fact that the bunch length at the solenoid actually increases with the initial bunch charge. Therefore the same argument applies as for the bunch charge.

From the prediction results and the ground truth a mean relative error was calculated over the whole test data set, yielding a mean prediction error for the emittance at the solenoid of \SI{1.0}{\percent}. Beam size and divergence at the solenoid are predicted very accurately with an error less than \SI{0.1}{\percent}. In addition to the beam parameters at the solenoid position, \textsc{ASTRA} input parameters were predicted. The laser spot size is predicted with an error down to \SI{0.7}{\percent}. Emission time and bunch charge are predicted with errors of \SI{32.1}{\percent} and \SI{27.1}{\percent} respectively. The fact that the laser spot size is predicted best out of the three input parameters is due to the fact that it has the strongest effect on the shape of the beam size vs.\ focusing strength data, especially for low charges (linear dependence of the thermal emittance). These results, as well as the percentage of predictions within a \SI{1}{\percent}, \SI{5}{\percent} and \SI{10}{\percent} relative error interval are summarized in Table~\ref{tab:pred_perf_table}.

\begin{table*}[htbp]
   \caption{Summary of the prediction performance for the different training label components. Mean error refers to the absolute prediction error and is calculated for the whole verification data set. The three remaining columns refer to the fraction of the verification data set within a certain prediction accuracy interval.}
   \begin{ruledtabular}
   \begin{tabular}{lcccc}
       & Mean Error (\SI{}{\percent}) & $<$ \SI{1}{\percent} Error (\SI{}{\percent}) & $<$ \SI{5}{\percent} Error (\SI{}{\percent}) & $<$ \SI{10}{\percent} Error (\SI{}{\percent})\\
      \hline
      Laser Spot Size & 0.71 & 76.59 & 99.84 & 100.00\\
      Emission Time & 32.14 & 2.46 & 11.67 & 21.05\\
      Bunch Charge & 27.05 & 10.24 & 52.74 & 76.74\\
      Emittance at the Solenoid & 1.01 & 55.67 & 99.81 & 100.00\\
      Beamsize at the Solenoid & 0.05 & 100.00 & 100.00 & 100.00\\
      Divergence at the Solenoid & 0.05 & 100.00 & 100.00 & 100.00\\
      Bunch Length at the Solenoid & 2.60 & 24.35 & 89.76 & 98.13\\
   \end{tabular}
   \end{ruledtabular}
   \label{tab:pred_perf_table}
\end{table*}

\begin{figure}[htbp]
  \centering
  \includegraphics[width=\columnwidth]{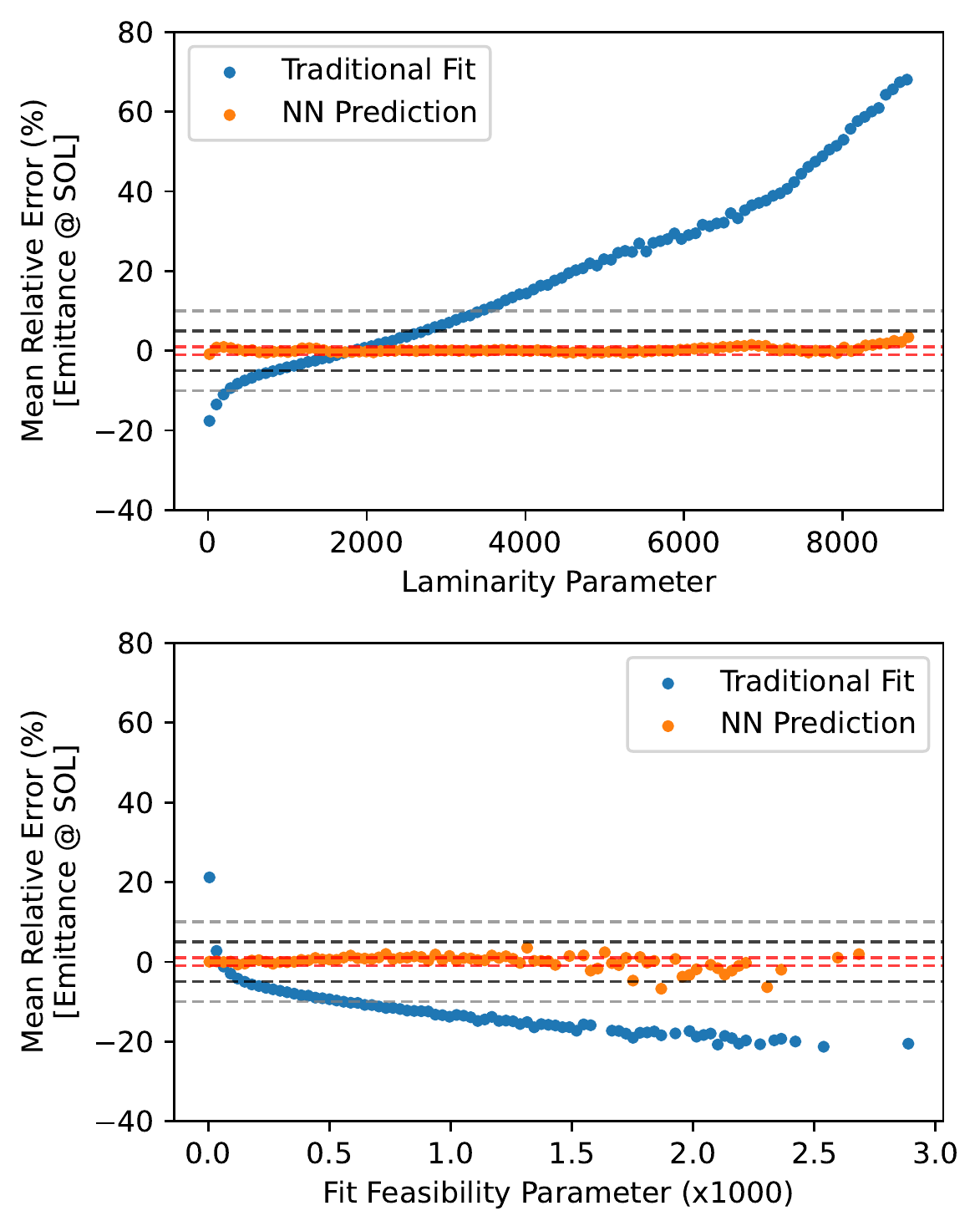}
  \caption{Mean emittance FCNN prediction performance, as well as traditional fit results vs.\ the laminarity parameter (top) and the fit feasibility criterion (bottom). The data is calculated for 100 equally sized laminarity parameter slices. $\gamma = 6.8$ for this population. \emph{Note that the bottom plot shows the fit criterion values multiplied by 1000 for readability.}}
  \label{fig:error_perf_comp_Ideal}
\end{figure}

The main goal of the study is to find a better way to determine the transverse emittance from phase advance scan data in the $\rho \gg 1$ regime, as well as in regimes where Eq.~\ref{eq:fit_criterion_math} does not hold. It is hence useful to evaluate the emittance prediction performance in form of the relative error versus these two quantities. Figure~\ref{fig:error_perf_comp_Ideal} shows the result of this analysis using both the FCNN, as well as the traditional fit routine (cf. Sec.~\ref{sec:MeasurementTechnique}). 

As expected from Fig.~\ref{fig:space_charge_dominance_train_data_sets} and Fig.~\ref{fig:feasibility_train_data_sets}, the traditional fit yields inaccurate results across the whole data set. The FCNN, on the other hand, performs much better even for very high values of $\rho$.
\section{Extension to Measured Data}
So far, the neural network was trained and tested solely with ideal phase advance scan data. This means that the training, validation, as well as the test data sets contain only perfectly evenly spaced data points with flawless beam size values. In addition, all scans were simulated within the same focusing strength range. Although this approach yields very good prediction performance for simulated data, it cannot be applied to experimental data, for several reasons. First, in a measured data set it is not guaranteed that all data points are evenly spaced. It is also not guaranteed that the scan range corresponds to the trained one and that the measured values of the focusing strength are correct \footnote{This issue is already somewhat alleviated by using $\delta B_i$, as described in Sec.~\ref{sec:Methodology}}. Finally, the measured beam sizes are subject to jitter and systematic measurement errors like resolution limitations.

\begin{figure}[htbp]
  \centering
  \includegraphics[width=\columnwidth]{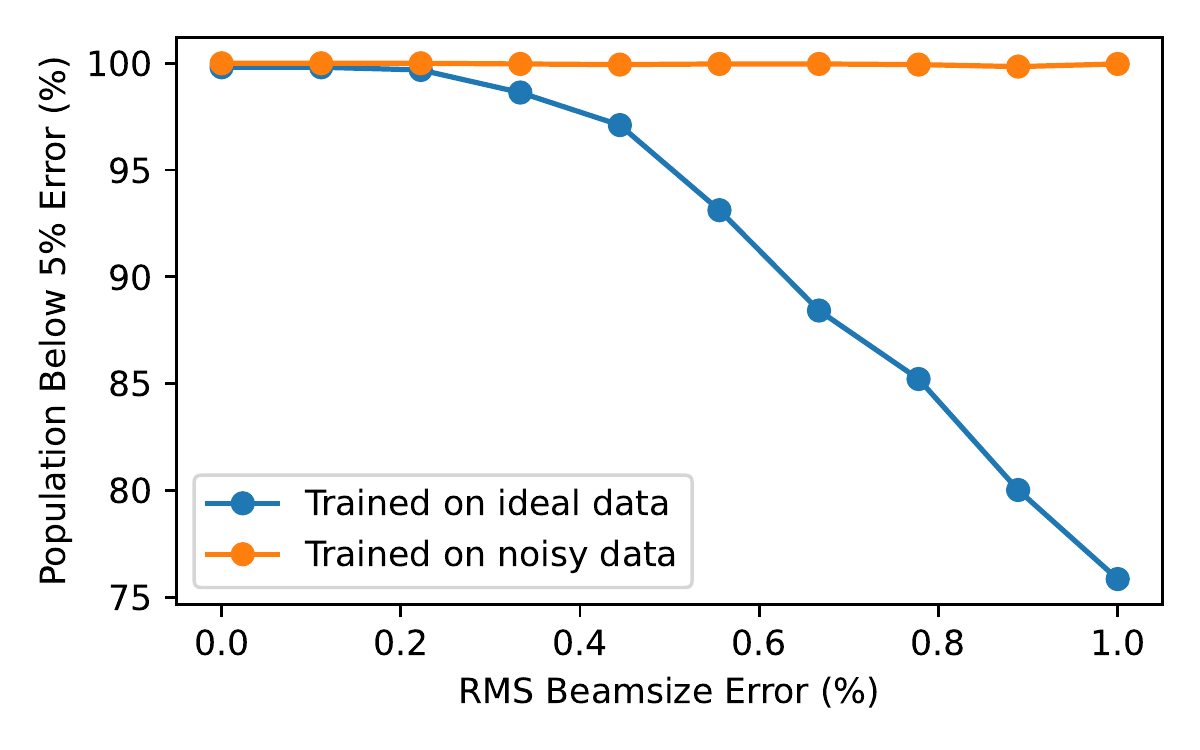}
  \caption{Population with less than \SI{5}{\percent} prediction error on the emittance vs.\ different artificial noise levels. The FCNN layout is the same in both shown cases. \emph{Note that in order to enable a direct comparison, the FCNNs used here were trained on the data sets with fixed scan range and constant spacing. This explains the generally better performance compared to Table~\ref{tab:pred_perf_table_2}.}}
  \label{fig:noise_performance}
\end{figure}

In order to take all of this into account, the training procedure was modified. Each data set is now created with a slightly different focusing strength scan range. Since the number of scan points is kept constant, the spacing between data points is now slightly different every time, which might also be the case in reality. Furthermore, the first and last focusing setting were added to the range of predicted parameters ($\rightarrow M_\text{out} = 9$). Measurement errors were taken into account by generating noisy data sets from the ideal sets by adding normally distributed errors. Both relative and absolute errors on focusing strength and beam size were considered with magnitudes based on experience at ARES. From each data set, $N_\text{err} = 100$ noisy sets with relative errors and $N_\text{err}$ noisy sets with absolute errors were generated. In addition, the procedure was repeated, this time enforcing a \SI{10}{\micro\meter} resolution limit on the beam sizes. Including the ideal data, the total number of data sets now increases to $N_\text{tot} = 2(2N_\text{err}+1)\cdot N = 6458532$. To visualize the importance of training the FCNN with noisy data, we fed data sets with increasing artificial noise to both the network trained solely on ideal data, as well as one trained on noisy data. The results are shown in Fig.~\ref{fig:noise_performance}. It can be seen that the population with less than \SI{5}{\percent} decreases significantly with noise level, if the FCNN is not trained on noisy data.

We performed the same general analysis for the new network, as described above, and saw the same overall behaviour. The results are summarized in Table~\ref{tab:pred_perf_table_2}. Compared to the network based on ideal data, the performance is slightly worse, but still for the majority of data sets the emittance is predicted with less than $\SI{5}{\percent}$ error.

\begin{table*}[htbp]
   \caption{Summary of the prediction performance of the modified network for the different training label components. Mean error refers to the absolute prediction error and is calculated for the whole verification data set. The three remaining columns refer to the fraction of the verification data set within a certain prediction accuracy interval.}
   \begin{ruledtabular}
   \begin{tabular}{lcccc}
       & Mean Error (\SI{}{\percent}) & $<$ \SI{1}{\percent} Error (\SI{}{\percent}) & $<$ \SI{5}{\percent} Error (\SI{}{\percent}) & $<$ \SI{10}{\percent} Error (\SI{}{\percent})\\
      \hline
      Laser Spot Size & 2.76 & 22.25 & 87.85 & 98.83\\
      Emission Time & 15.97 & 4.25 & 21.08 & 43.48\\
      Bunch Charge & 14.13 & 6.44 & 30.89 & 67.20\\
      Emittance at the Solenoid & 2.51 & 29.43 & 89.17 & 97.51\\
      Beamsize at the Solenoid & 0.38 & 96.34 & 100.00 & 100.00\\
      Divergence at the Solenoid & 0.38 & 96.63 & 100.00 & 100.00\\
      Bunch Length at the Solenoid & 4.48 & 12.74 & 60.03 & 95.02\\
   \end{tabular}
   \end{ruledtabular}
   \label{tab:pred_perf_table_2}
\end{table*}

\section{Measurements at the ARES linac}
As a real world test, we conducted emittance measurements using the phase advance scan technique at the ARES linac at DESY. The layout of the measurement setup is shown in Fig.~\ref{fig:ARES_sketch}. We took data for several bunch charges by adjusting an attenuator in the cathode laser beamline. The charge was measured both with a Faraday cup, which can be inserted into the beamline instead of the scintillating screen and a cavity based charge monitor $\sim \SI{0.7}{\meter}$ downstream of the screen \cite{Lipka:166172}. All measurements shown here were performed according to the procedure introduced in Sec.~\ref{sec:MeasurementTechnique}. Transverse beam sizes were determined from camera images of a scintillating Ce:GAGG (Cerium doped Gadolinium Aluminium Gallium Garnet) screen. The spatial resolution of the system is specified to be $\sim \SI{10}{\micro\meter}$ \cite{Wiebers:166567}.

\begin{figure}[htbp]
  \centering
  \includegraphics[width=\columnwidth]{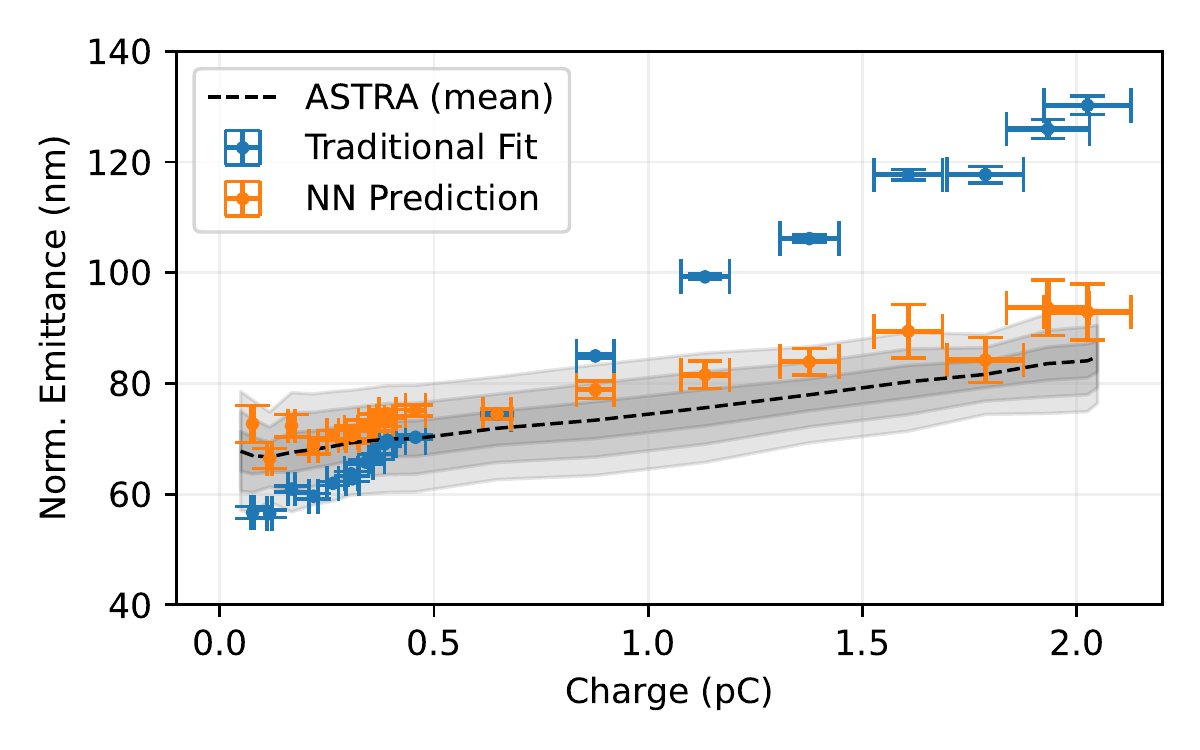}
  \caption{Emittance measurements conducted at the ARES linac at DESY using the phase advance scan technique for different bunch charges. The black dashed line denotes the mean ASTRA simulation result. The shaded areas correspond to the 1, 2 and 3\,$\sigma$ uncertainty of the simulation, based on 100 simulations with normally distributed input parameters.}
  \label{fig:emittance_measurement_ARES}
\end{figure}

Figure~\ref{fig:emittance_measurement_ARES} shows the emittance values obtained from the measurements using both the FCNN, as well as the traditional fit method. The data points are compared to an ASTRA simulation including space charge based on the machine settings at the day of the measurements, including uncertainty. It can be seen that the FCNN results are much closer to the expected values than the results obtained from the envelope equation fit. It is interesting to note that Fig.~\ref{fig:emittance_measurement_ARES} reproduces the expected behaviour shown in Fig.~\ref{fig:error_perf_comp_Ideal}, as the fit underestimates the emittance for charges $<\SI{0.5}{\pico\coulomb}$ and overestimates them for higher charges. The FCNN result follows the ASTRA curve much closer, mostly staying within the uncertainty of the simulation.

\begin{figure}[htbp]
  \centering
  \includegraphics[width=\columnwidth]{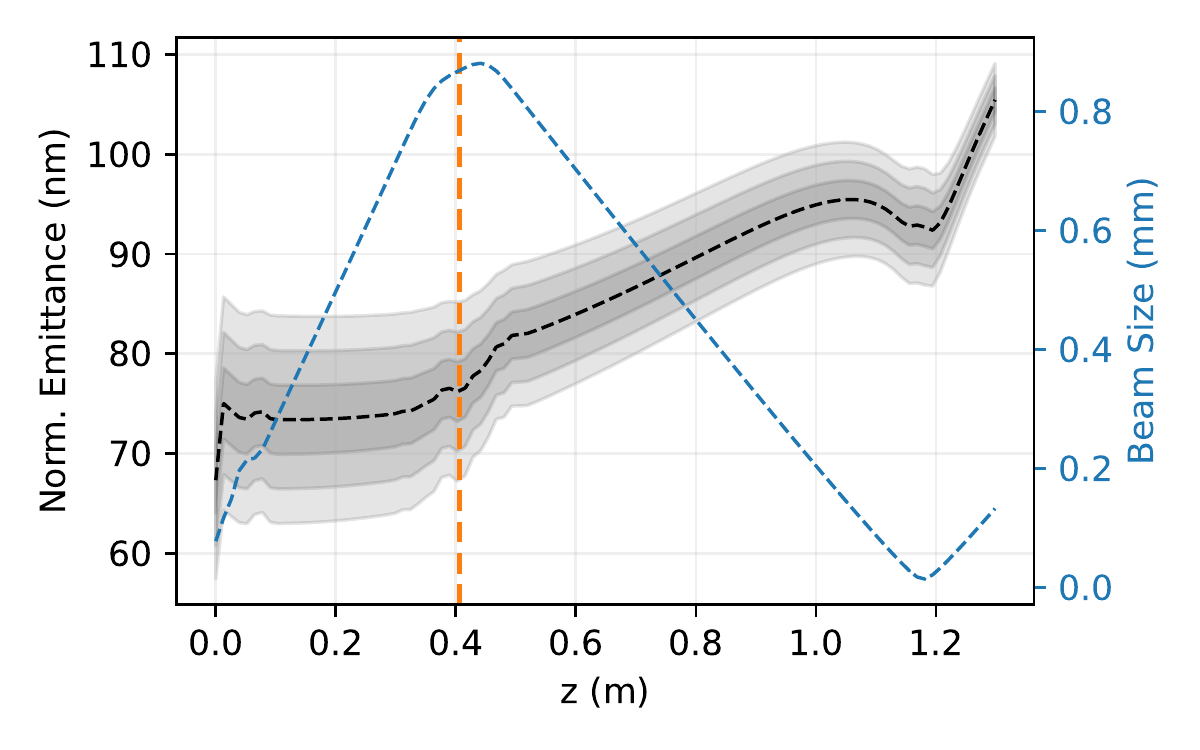}
  \caption{ASTRA simulation of the evolution of the transverse normalized emittance between gun and grid for a \SI{1}{\pico\coulomb} bunch with $\gamma = 6.8$. The orange dashed line marks the position of the solenoid, which is used to perform the phase advance scan measurements. The black dashed line denotes the mean ASTRA simulation result. The shaded areas correspond to the 1, 2 and 3\,$\sigma$ uncertainty of the simulation, based on 100 simulations with normally distributed input parameters. The blue dashed line shows the evolution of the beam size – the beam reaches its focus slightly before the grid.}
  \label{fig:emittance_measurement_ARES_grid_sim}
\end{figure}

In order to cross-check the obtained results, we performed additional emittance measurements using a grid mask based method, as described in \cite{PhysRevAccelBeams.21.102802}, using the same machine setup. Three different grids were used for the measurements: An SPI G200TH TEM grid \footnote{SPI G200TH TEM Grid, available at \url{https://www.2spi.com/item/202hc-xa}, \emph{last access: 22nd April 2022}}, an SPI G300 TEM grid \footnote{SPI G300 TEM Grid, available at \url{https://www.2spi.com/item/2030c-xa}, \emph{last access: 22nd April 2022}} and a custom made pepper pot \footnote{Custom made Tungsten pepper pot (laser drilled). Hole diameter: \SI{15}{\micro\meter}, pitch: \SI{85}{\micro\meter}, thickness: \SI{20}{\micro\meter}}. Measurements were performed for two different charge settings, \SI{1}{\pico\coulomb} and \SI{2}{\pico\coulomb}. These values were deliberately chosen to be in the strongly space charge dominated regime, where the traditional fit yields particularly bad results. At ARES, the grid masks are installed at the same $z$-position as the screen used to record the phase advance scan data. This means that the emittance obtained from the grid measurements will always be different from the phase advance scan result, as the phase advance scan yields the emittance at the position of the focusing element. The emittance is furthermore expected to be different, because in order to image the grid, the beam needs to be focused slightly before the grid, which can lead to emittance growth. Nevertheless, it is still possible to compare the measured values to ASTRA simulations, which would show that the ARES setup depicted in Fig.~\ref{fig:ARES_sketch} can be well simulated with ASTRA. This would validate the FCNN results indirectly. Figure~\ref{fig:emittance_measurement_ARES_grid_sim} shows an ASTRA simulation of the grid measurement using a \SI{1}{\pico\coulomb} bunch with $\gamma = 6.8$. It can be seen that the emittance is strongly affected by focusing the beam down. The measurement results from all three grids, as well as the expected values from the ASTRA simulation are summarized in Table~\ref{tab:grid_meas_comp}. The results are very close to the expected value in the \SI{1}{\pico\coulomb} case. The measurement of the \SI{2}{\pico\coulomb} beam shows a slightly higher than expected emittance value, which is in line with the high uncertainty of the high charge results shown in Fig.~\ref{fig:emittance_measurement_ARES}. We hence conclude that ASTRA simulates the ARES beamline shown in Fig.~\ref{fig:ARES_sketch} well.
\begin{table}[htbp]
   \caption{Summary of the grid based measurements, used to verify the FCNN results. The measurements are compared to the ASTRA simulation result.}
   \begin{ruledtabular}
   \begin{tabular}{lcc}
      Method & $\varepsilon _\text{n}$ (\SI{}{\nano\meter}) @ \SI{1}{\pico\coulomb} & $\varepsilon _\text{n}$ (\SI{}{\nano\meter}) @ \SI{2}{\pico\coulomb}\\
      \hline
      \textbf{ASTRA Simulation} & $\mathbf{105.5 \pm 1.2}$ & $\mathbf{152.9 \pm 3.0}$ \\
      \hline
      SPI G200TH TEM Grid & $107.4 \pm 5.8$ & $165.4 \pm 15.7$ \\
      SPI G300 TEM Grid & $102.7 \pm 3.8$ & $156.0 \pm 15.6$ \\
      Pepper Pot & $105.6 \pm 11.9$ & $156.1 \pm 27.4$ \\
      \textbf{Mean (Measurements)} & $\mathbf{105.2 \pm 8.0}$ & $\mathbf{160.5 \pm 20.3}$
   \end{tabular}
   \end{ruledtabular}
   \label{tab:grid_meas_comp}
\end{table}
\subsection{Prediction of other parameters}
As discussed above, the FCNN also predicts fixed machine parameters and other charge dependent beam parameters to varying accuracy (see Table~\ref{tab:pred_perf_table_2}). Table~\ref{tab:pred_fixed_params} summarizes the predicted fixed machine parameters in comparison to what was used in the experiment. It can be seen that both the mean laser spot size and pulse length are predicted to be larger than the expected values. Since measuring the laser spot size on the cathode directly is very difficult in the ARES setup, the predicted values fall within the uncertainty. As discussed in Sec.~\ref{sec:ResultsAndComparison}, the prediction of the laser pulse length should be treated with caution. The solenoid scan range is predicted well within the uncertainties.
\begin{table}[htbp]
   \caption{Prediction results for fixed machine parameters.}
   \begin{ruledtabular}
   \begin{tabular}{lcc}
      Parameter & Prediction & Experiment\\
      \hline
      Laser Spot Size (\SI{}{\micro\meter}) & $338.5 \pm 0.5$ & $320 \pm 30$ \\
      Laser Pulse Length (\SI{}{\femto\second}, rms) & $87.34 \pm 0.04$ & $76 \pm 8$ \\
      Solenoid Field -- Start (mT) & $130.516 \pm 0.001$ & $130.5 \pm 0.1$ \\
      Solenoid Field -- End (mT) & $150.783 \pm 0.001$ & $150.7 \pm 0.1$ \\
   \end{tabular}
   \end{ruledtabular}
   \label{tab:pred_fixed_params}
\end{table}

\begin{figure}[htbp]
  \centering
  \includegraphics[width=\columnwidth]{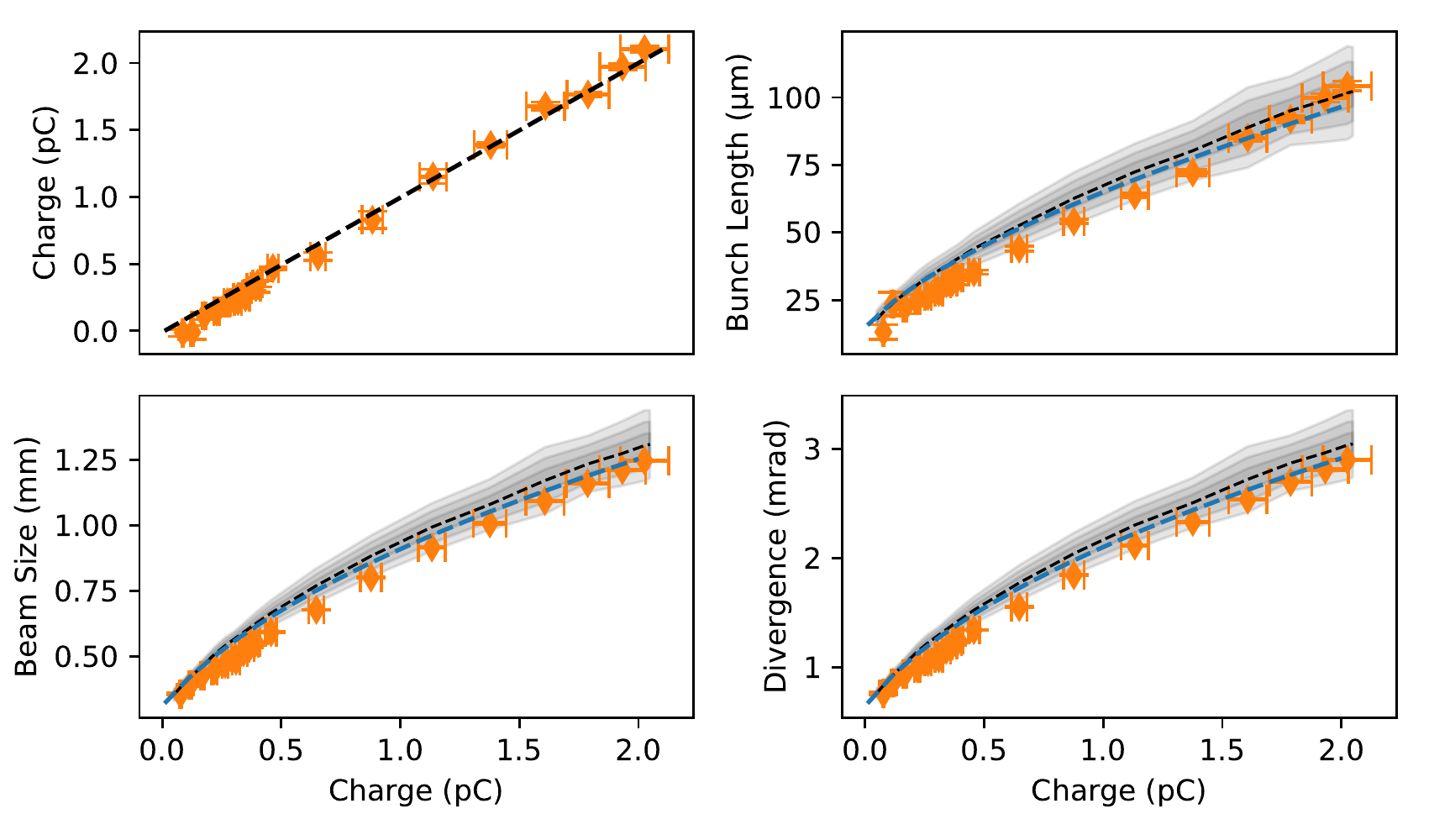}
  \caption{Prediction results for other charge dependent beam parameters. The black dashed line denotes the mean ASTRA simulation result. The shaded areas correspond to the 1, 2 and 3\,$\sigma$ uncertainty of the simulation, based on 100 simulations with normally distributed input parameters. The blue dashed line corresponds to an ASTRA simulation using the mean predicted laser spot size and pulse length (see Table~\ref{tab:pred_fixed_params}).}
  \label{fig:pred_variable_params}
\end{figure}

Figure~\ref{fig:pred_variable_params} shows the prediction results for the beam charge, as well as the charge dependent beam parameters \emph{bunch length}, \emph{beam size} and \emph{beam divergence} at the solenoid. As in Fig.~\ref{fig:emittance_measurement_ARES}, the data points are compared to an ASTRA simulation including space charge based on the machine settings at the day of the measurements, including uncertainty. It can be seen that the beam charge prediction fits the measured values well. Both beam size and beam divergence follow the ASTRA curve well, albeit at the lower end of the uncertainty, denoted by the shaded area. The bunch length follows a more linear charge dependence than expected from the ASTRA simulation, which might be attributed to either the not fully known temporal and spatial laser pulse shape at the cathode, as well as the overall prediction performance of parameters of the longitudinal phase space (see Sec.~\ref{sec:ResultsAndComparison}). 

In order to cross-validate the prediction results, an ASTRA simulation using the mean predicted laser spot size and pulse length (see Table~\ref{tab:pred_fixed_params}) was performed. The results are shown in Fig.~\ref{fig:pred_variable_params} as the blue dashed line. Indeed, a larger spot size and longer pulse length lead to results closer to the lower end of the uncertainty in all three cases, which can be explained by the reduced charge density. Remaining discrepancies might be explained by the not fully known temporal and spatial laser pulse shape at the cathode.
\section{Conclusion and Outlook}
We have shown in simulation that a pre-trained fully connected neural network can be used to predict the transverse emittance from phase advance scan data even in the $\rho \gg 1$ regime and in case a traditional envelope equation based fit is mathematically not feasible. We have optimized the network for real-world measurement data and achieved $< \SI{5}{\percent}$ error for the majority of the test data set population (\SI{89.2}{\percent}), resulting in a mean relative error of \SI{2.5}{\percent}. We have applied our method to measurements conducted at the ARES linac at DESY and compared the predictions to numerical simulations using the well benchmarked code ASTRA, as well as results obtained from the traditional fit method. As expected from the simulation study, the FCNN predictions are much closer to what is expected from the numerical simulation. We have furthermore cross-validated the results using additional emittance measurements based on a grid mask based method.

In addition to the transverse emittance, the network also predicts other key beam and machine parameters to varying accuracy. While quantities directly tied to the transverse phase space are predicted as accurate or better than the emittance, quantities tied to the longitudinal phase space, such as the bunch length, are predicted less accurate, as expected. It should be noted, that in our study the gun setting is not a variable in the process of training the FCNN. This means that for each gun setting (gradient and phase) a separate FCNN has to be trained. Inclusion of these two parameters could be part of a future study. Furthermore, difficult to directly access parameters, such as the thermal emittance could be added. In conclusion, we have demonstrated that pre-trained FCNNs can be a powerful tool for the analysis of previously difficult to interpret data sets.
\section{Model Availability}
The FCNN models are available from the corresponding author upon request in TensorFlow format.
\begin{acknowledgments}
We acknowledge support from DESY (Hamburg, Germany), a member of the Helmholtz Association HGF. Specifically, we would like to thank the DESY beam diagnostics group for their support during the measurement runs at ARES. We also thank the DESY Maxwell team for providing the compute resources we used to generate the training data sets in a reasonable time. Finally, we thank R.~Mayet (Halodi Robotics AS) for NN-related consulting.
\end{acknowledgments}
\bibliography{NeuralNetSolenoidScan}
\end{document}